\documentclass[aps,prb,twocolumn]{revtex4}
\usepackage{graphicx}

\begin{document}

\title{Characterization of an entangled system of two superconducting qubits using a multiplexed capacitance measurement}

\author{M. D. Shaw,$^1$ J. F. Schneiderman,$^{1,2}$ J. Bueno,$^3$ B. S. Palmer,$^4$ P. Delsing,$^{2,3}$ and P. M. Echternach$^{1,3}$}

\address{$^1$ Department of Physics and Astronomy, University of Southern California, Los Angeles, California 90089, USA}
\address{$^2$ Microtechnology and Nanoscience, MC2, Chalmers University of Technology, 412 96 G\"{o}teborg, Sweden}
\address{$^3$ Jet Propulsion Laboratory, California Institute of Technology, Pasadena, California 91109, USA}
\address{$^4$ Laboratory for Physical Sciences, College Park, Maryland 20740, USA}

\date{\today}

\begin{abstract}
We characterize a pair of Cooper-pair boxes coupled with a fixed capacitor using spectroscopy and measurements of the ground-state quantum capacitance. We use the extracted parameters to estimate the concurrence, or degree of entanglement between the two qubits. We also present a thorough demonstration of a multiplexed quantum capacitance measurement technique, which is in principle scalable to a large array of superconducting qubits. 
\end{abstract}

\maketitle

\section{Introduction}

Entanglement is a unique feature of quantum mechanics which is of great fundamental interest, and is also an essential resource for quantum information processing. In the field of superconducting electronics, macroscopic entangled states of two or more qubits have recently been investigated in charge,\cite{Nakamura1,Nakamura2qubit,Yale2qubit} flux,\cite{Mooij,Clarke,Ilichev} and phase\cite{Wellstood,Martinis1,Simmonds} qubits. The pioneering experiments with fixed coupling explored the two-qubit state space through coherent oscillations\cite{Nakamura1} and microwave spectroscopy,\cite{Wellstood} mapping out the avoided level crossings characteristic of coupled qubits. Other key developments from the perspective of quantum information processing were the demonstration of controlled gate operations\cite{Nakamura2qubit} and simultaneous single-shot two-qubit measurements\cite{Martinis1} in the case of fixed coupling. More recently, the strong-coupling limit of circuit QED has been used to exchange quantum information between two qubits via a superconducting resonator,\cite{Yale2qubit,Simmonds} and a variety of tunable coupling schemes have been demonstrated experimentally.\cite{Clarke,JenaCoupling,Nakamura3} Furthermore, extensive work has been performed to investigate the ground state entanglement of systems of two, three, and four coupled qubits in the context of adiabatic quantum computing,\cite{Ilichev,JenaCoupling,Jena4qubit} and such measurements have been directly compared with spectroscopic characterization and Landau-Zener interferometry.\cite{IlichevRecent}

In this work, we characterize a system composed of two single Cooper-pair box (SCB) charge qubits, coupled with a fixed capacitor, using a frequency-multiplexed quantum capacitance measurement technique. The qubit parameters, including the coupling energy between the two qubits, are extracted spectroscopically and via measurements of the ground state capacitance. These parameters can then be used to estimate the degree of entanglement for the two-qubit system as a function of gate charge, although such an inference does not in itself constitute a direct observation of the entangled state, as can be established using state tomography.\cite{Martinis}

We also present the multiplexed quantum capacitance measurement (QCM), which is an effective method for probing the state of multiple qubits with a single RF line. In a single qubit, the QCM is a dispersive measurement of the reactive response of an LC oscillator coupled capacitively to the qubit island.\cite{DelsingQCR,FinnQCR} The capacitance of the qubit has, in addition to its geometric capacitance, a term which is determined by the second derivative of the qubit energy with respect to the gate charge, so that the overall capacitance of the oscillator depends on the qubit state. The state of the SCB can be obtained in a capacitance measurement by monitoring the center frequency of the oscillator with RF reflectometry. The oscillator is tuned to a frequency much lower than the qubit energy level splitting, minimizing measurement back action and disturbance to the qubit. This also has the benefit of filtering high-frequency noise from the RF line. A primary advantage of the QCM technique as compared to some earlier measurement devices, such as the single-electron transistor, is its applicability for qubit state discrimination at the degeneracy point, where dephasing due to low-frequency voltage fluctuations is minimized.\cite{Vion} In principle, the QCM technique can be used to perform quantum-limited measurements.\cite{QuantumLimitedQCR} 

As superconducting circuits grow in complexity, efficient multiplexing schemes are required for practical operation at millikelvin temperatures. By coupling each qubit to a high-Q oscillator of a different frequency, an array of qubits can be efficiently read out with multiplexed QCM by applying a frequency comb to a single RF line. In this work, we demonstrate the use of such a technique to characterize a system of two coupled qubits. A similar concept is also being used to read out arrays of superconducting radiation detectors.\cite{MKID} 

  In the four-level approximation, the Hamiltonian for a two-qubit system coupled by a fixed capacitance $C_m$ is given by\cite{Nakamura2qubit,Storcz}

\begin{eqnarray}
\label{Hamiltonian}
H = -\frac{1}{2}\biggl\{\left[ 4E_{C1} \left( \frac{1}{2} - n_{g1} \right) + 2E_m \left( \frac{1}{2} - n_{g2} \right)\right] \sigma_{z1} \nonumber \\
+ \left[ 4E_{C2} \left( \frac{1}{2} - n_{g2} \right) + 2E_m \left( \frac{1}{2} - n_{g1} \right)\right] \sigma_{z2} \nonumber \\
+ E_{J1}\sigma_{x1} + E_{J2}\sigma_{x2} - 2E_m \sigma_{zz} \biggr\}
\end{eqnarray}

\noindent where $E_{Cj} = \frac{e^2 C_{\Sigma (3-j)}}{2\left( C_{\Sigma1}C_{\Sigma2} - C_m^2 \right)}$ and $E_{Jj}$ are the charging and Josephson energies, respectively, for the $j$th qubit, $E_m = \frac{e^2 C_{m}}{C_{\Sigma1}C_{\Sigma2} - C_m^2}$ is the mutual coupling energy between the two qubits,  and $C_{\Sigma j}$ is the total capacitance for the $j$th qubit. In Eqn.~(\ref{Hamiltonian}), $\sigma_{x,z}$ are the standard Pauli matrices, $\sigma_{z1} = \sigma_z \otimes 1$, $\sigma_{z2} = 1 \otimes \sigma_z$, and $\sigma_{zz} = \sigma_z \otimes \sigma_z$.  Note that the interqubit coupling is along the $zz$ axis, and is not controllable with the gate voltages. The coupling between qubits introduces avoided level crossings and a gate voltage asymmetry into the energy spectrum.

In section II, we discuss the multiplexed QCM measurement, while in the remainder of the paper we present a characterization of the entangled two-qubit system. Section IIa provides a short review of the QCM concept for a single qubit, while section IIb presents the details of the experiment and a discussion of device fabrication. In section IIc, we qualitatively discuss the performance of the lumped-element superconducting oscillators used in the measurement, while in section IId we present a detailed demonstration of the multiplexed QCM technique. In section IIIa, we present a characterization of the two-qubit system, where we estimate the qubit parameters and coupling energy using microwave spectroscopy and an analysis of the ground-state capacitance. In section IIIb, we use these parameters to estimate the ground-state concurrence, and discuss entanglement in the two-qubit system. 

\section{Multiplexed Capacitance Measurement}
\subsection{Overview}

Let us review the essential details of the quantum capacitance measurement for one qubit. For a single parallel lumped-element LC tank circuit coupled capacitively to a single qubit and transmission line, the overall capacitance of the oscillator when the qubit is in its \it i\rm th energy eigenstate is

\begin{equation}
\label{capacitance}
C_i = C_T + C_C + \frac{C_{RF} C_J}{C_{RF} + C_J} - \frac{C_{RF}^2}{4e^2}\frac{\partial^2 E_i}{\partial n_g^2}
\end{equation}

\noindent where $C_T$ is the tank circuit capacitance, $C_C$ is the coupling capacitance between the tank circuit and the 50 $\Omega$ transmission line, $C_{RF}$ is the RF gate capacitor, and $C_J$ is the combined qubit junction capacitance, $E_i$ is the \it i\rm th eigenvalue of the qubit, and $n_g = C_{g} V_g / 2e$ is the normalized qubit gate charge, where $C_{g}$ is the control gate capacitance. The fourth term in Eq.~(\ref{capacitance}) is referred to as the quantum capacitance, and is proportional to the curvature of the qubit energy level. For a single qubit in the two-level approximation, where $\alpha \equiv E_J/4E_C \ll 1$, the quantum capacitance in the ground $(+)$ and first excited $(-)$ states is given by

\begin{equation}
\label{QConequbit}
C_Q^\pm = \pm \frac{C_{RF}^2}{C_\Sigma}\frac{\alpha^2}{\left( (1-2n_g)^2 + \alpha^2 \right)^{3/2}}
\end{equation}

\noindent where $C_\Sigma$ is the total capacitance of the qubit island. For the coupled two-qubit system, the eigenstates are best evaluated numerically, but the quantum capacitance will approximately reduce to Eq.~(\ref{QConequbit}) when one of the two qubits is far from its degeneracy point. By measuring the phase shift of a reflected RF signal, one can extract the quantum capacitance, as described in section IId. Since the ground and first excited states have opposite curvature at the degeneracy point, this technique can be used to measure the state of the qubit directly at its operating point. 

\subsection{Experimental Design and Setup} 
 
Next we describe a multiplexed version of QCM for two charge qubits based on the single Cooper-pair box (SCB). Two parallel lumped-element tank circuits with different inductances are capacitively coupled to a single transmission line, which is probed with a two-tone RF signal. The reflected signal is demodulated in a homodyne technique with two analog quadrature mixers. By measuring the phase shift of each signal, we may simultaneously measure the quantum capacitance of each qubit. Such multiplexing schemes have been previously employed with the RF single-electron transistor,\cite{MultiplexedSET,DiffSET} and can be readily scaled to read out a large array of qubits by using high-Q oscillators. In this experiment, the Q of the lumped-element tank circuits is of order 1000.  
		
\begin{figure}
\centering
\includegraphics{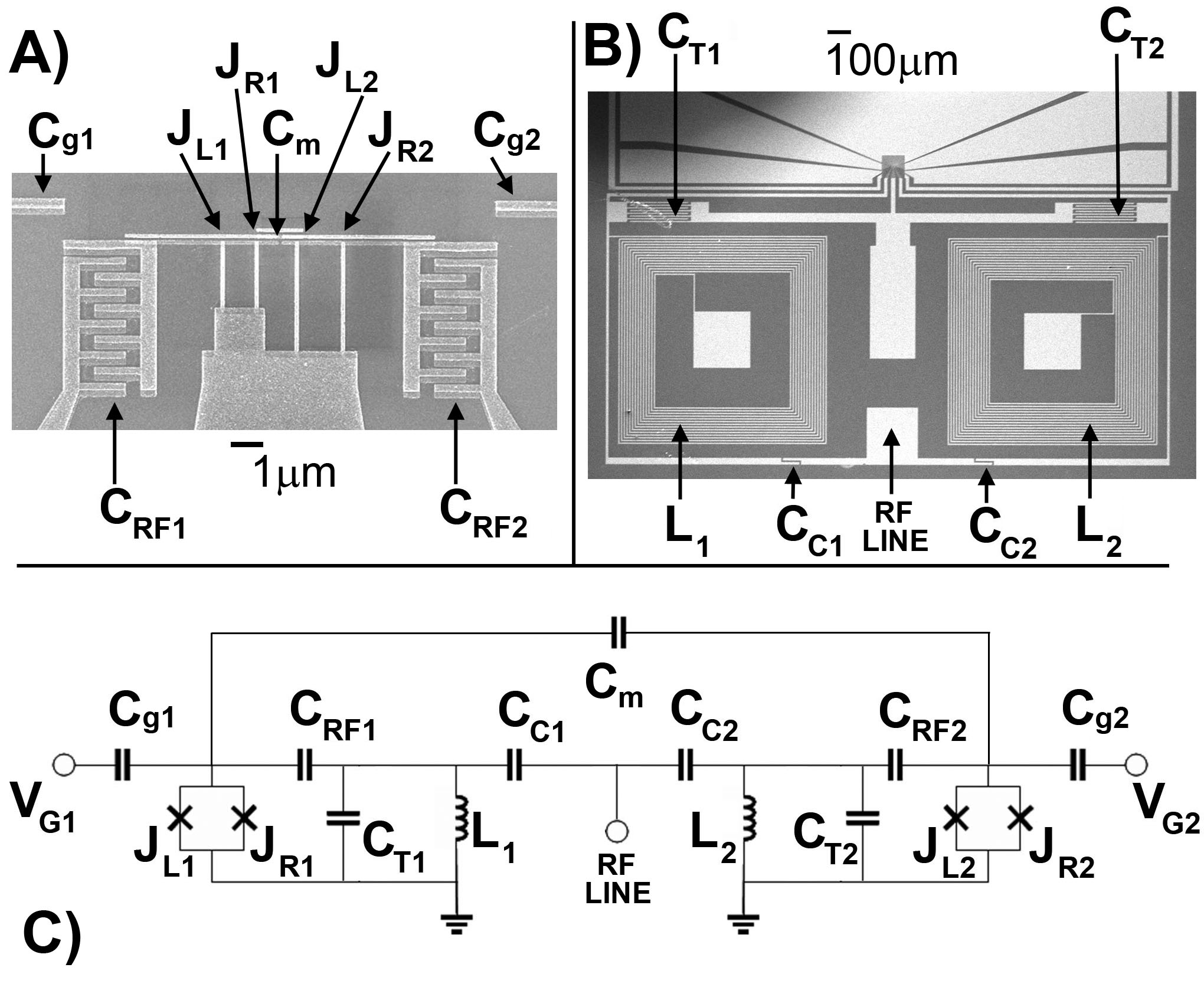}
\caption{\bf A)\rm~SEM image of the qubit structures for sample 1. Fabrication is performed using electron-beam lithography and double-angle evaporation. The interdigitated capacitors in the center are the RF gates, while the short leads to the sides are the control gates. The qubit islands are the thin strips at the center, and include one side of each interdigitated capacitor. The two qubits are coupled with a fixed coupling capacitor formed by the shadows of the qubit islands, which are spanned by a small bridging structure. \bf B)\rm ~SEM image of the optically patterned tank circuits for sample 1. The left spiral inductor has 15 turns ($L_1 = 380$ nH) while the right has 17 turns ($L_2 = 430$ nH). \bf C)\rm ~Circuit diagram showing qubit structures and tank circuits.}
\end{figure}
	
The circuit layout and experimental setup is shown in Fig.~1. The qubit devices were fabricated using a conventional shadow-mask evaporation technique. The qubit features shown in Fig.~1A were patterned with electron-beam lithography, and the qubits themselves each consist of a pair of small-area Al/AlOx/Al tunnel junctions in a DC-SQUID configuration. Note that the areas of the two loops are different, so that the Josephson energies of the two qubits can be tuned concurrently with a single superconducting magnet. 

The two qubits are coupled by a fixed capacitor, which is formed from the shadow of the islands themselves, spanned by a small metal bridge. While the bridge effectively introduces a pair of tunnel junctions within the coupling capacitor itself, the area of these junctions is large enough (720 $\times$ 230 nm) that the single Cooper-pair charging energy of the bridge is negligible. Assuming that the junction capacitance scales linearly with the area, we estimate this charging energy of the bridge to be 19 mK. The tank circuits are coupled directly to the qubit islands with large interdigitated capacitors. 

The parallel LC tank circuits, which are shown in Fig.~1B, are patterned with photolithography and made from a superconducting Al/Ti/Au trilayer with respective thicknesses of 300, 200, and 200 \AA. This material acts as a superconductor with a suppressed transition temperature $T_C$ = 450 mK,\cite{leduc} which acts as a trap for quasiparticles in the leads.  Fig.~1C shows the circuit layout. Significant high-frequency filtering of the RF signal is performed with a CuNi microcoaxial transmission line thermally anchored to each refrigeration stage, and a cold circulator mounted to the mixing chamber is used to isolate the reflected signal return path from the feedline. The resonant frequencies of the right and left tank circuits are 545 and 588 MHz, respectively, which are an order of magnitude smaller than the qubit energy level spacing. 

In Fig.~1C, the SCB junctions for qubits 1 and 2 are labeled $J_{R1,2}$ and $J_{L1,2}$. The qubit islands are capacitively coupled to the LC oscillators through the RF gate capacitors, labeled $C_{RF1,2}$. These RF gates are the large interdigitated capacitors shown in Fig.~1A. The oscillators themselves are lumped-element parallel LC tank circuits, the components of which are labeled $L_{1,2}$ and $C_{T1,2}$ in Fig.~1C. Finally, the tank circuits are coupled to the 50~$\Omega$ transmission line through RF coupling capacitors $C_{C1}$ and $C_{C2}$. Nominal tank circuit parameters are $L_1 = 380$ nH, $L_2 = 430$ nH, $C_{T1} = 172$ fF, $C_{T2} = 177$ fF, and $C_{C1} = C_{C2} = 20$ fF. 

The quality factor Q of the oscillators are on the order of 1000, as opposed to $\sim20$ for typical implementations of the RF-SET and previous QCM measurements. This higher Q value is due to very small coupling capacitances between the LC oscillators and the 50~$\Omega$ transmission line. This results in a larger phase-shift of the reflected signal for a given quantum capacitance, thus permitting a more sensitive measurement and allowing the use of a smaller RF gate capacitor. This reduces the noise coupled from the RF line onto the qubit, as does the filtering effect of a high Q oscillator. Furthermore, a higher Q oscillator permits closer frequency spacing of the two oscillators, so that both resonances can fit within the 50 MHz bandwidth of the cold circulator. When the oscillators are monitored with RF reflectometry, the loss of signal bandwidth associated with a high Q oscillator limits the effectiveness of this circuit for short-pulse or single-shot readout. However, when the oscillators are allowed to self-oscillate by providing balanced feedback, a change in the oscillator frequency can in principle be detected in a time much shorter than the oscillator ringdown time.\cite{Rugar} Demodulation is performed by splitting the amplified RF output signal into two quadrature mixers. 

\subsection{Oscillator Performance}

In an ideal superconducting tank circuit, there is minimal intrinsic loss and the quality factor is set simply by the coupling capacitance to the 50 $\Omega$ transmission line.  However, in direct measurements of the linear response of the tank circuit as a function of temperature and magnetic field, we represent the intrinsic losses of the tank circuit at the resonance frequency with a series resistance of 1 $\Omega$ in devices fabricated on single-crystal quartz substrates. This value of the resistance was determined by fitting the tank circuit response to a lumped-element model. The series resistance manifests itself as a decreased amplitude and complex offset in the observed signal, which must be carefully accounted for when processing the data. 

For an applied RF power of -130 dBm, the current in the inductor is 0.4 nA, and this anomalous resistance corresponds to a power dissipation of roughly -160 dBm in the inductor. Measurements of the temperature and magnetic field dependence of the resonant response of similar tank circuits were consistent with predictions of a standard two-fluid model for the kinetic inductance and surface resistance of the superconducting trilayer. However, at low temperature and zero field, this model predicts negligible dissipation, so this series resistance cannot simply be ascribed to the AC surface resistance of the superconducting trilayer. Furthermore, an all-aluminum tank circuit fabricated with electron-beam lithography shows comparable losses.

Likewise, a straightforward estimate of the power dissipated into eddy currents in the Au-plated Cu sample box yields a dissipated power of -200 dBm, which is not enough to explain the loss. Another possible dissipation mechanism is electromagnetic loss in the substrate. Finite-element simulations predicted a small dissipative component in the resonant response given a material loss tangent typical of single-crystal quartz, but zero dissipation when given a loss tangent typical of sapphire. Similar devices fabricated on polished R-plane sapphire had an inductor series resistance of $0.20 \pm 0.08~\Omega$. However, the high dielectric constant of sapphire required a complete tank circuit redesign, with significantly less on-chip metal, so it is difficult to say with certainty whether the improvement is attributable to the improved loss characteristics of the substrate alone. All data shown in this paper was collected from devices using quartz substrates. While the tank circuit dissipation mechanism is still under investigation, significant improvement can be practically achieved simply by increasing the coupling capacitances $C_{C1,2}$ to the external transmission line. 

\subsection{Demonstration of Multiplexing}

A demonstration of the multiplexed QCM technique using sample 1 is shown in Fig.~2. In this experiment, tank circuits 1 and 2 are probed simultaneously, and the two qubit control gate voltages are ramped concurrently in the same direction. Equal voltage ramps from $\pm 3$ mV are applied to both qubit control gates, with a ramp frequency of 104.4 Hz. Note that in this configuration the system does not necessarily pass through the two-qubit mutual degeneracy point. In this experiment, the sample is mounted on the mixing chamber of a dilution refrigerator at its base temeperature of 20 mK. In this analysis, the capacitance signal is presumed to be 2e-periodic, i.e. there is negligible tunneling of non-equilibrium quasiparticles. When quasiparticles tunnel across the SCB junctions, the gate charge switches rapidly by one electron, and the time-averaged capacitance signal is a weighted average of the shifted and unshifted capacitance peaks. Such a signal is typically referred to as ``1e-periodic". The assumption that the observed signal is 2e-periodic at low temperatures is based on the observation of a transition to a 1e-periodic signal between 250 and 300 mK, which is consistent with the thermal occupation of equilibrium quasiparticle states. 

Figs.~2A-B show raw in-phase (I) and quadrature (Q) oscilloscope traces for both qubits, as labeled in the figure caption. These traces were recorded simultaneously. The x-axis displays the sweep time of the qubit gate ramp. Figs.~2C-D show the phase shift and magnitude response for both qubits, which represent the reactive and dissipative components of the qubit signal, respectively. As discussed above, the qubit response is transformed by the loss in the tank circuit, adding an offset in the complex plane. To compensate for this effect, the signal is recentered by subtracting this offset before computing the magnitude and phase shift. 

For qubit 1, the observed phase shift is $171^\circ$, while for qubit 2 the phase shift is $117^\circ$. The difference in the overall magintude of the phase shifts for the two qubits is due to the difference in $E_{J1,2}$ for this particular value of the DC flux bias, which was tuned to maximaize the phase shift for qubit 1. The observed phase shifts are quite large compared to previous QCM experiments, which were typically less than $10^\circ$.\cite{DelsingQCR,FinnQCR} This is a result of the higher Q value for the oscillators and the large RF gate capacitances. In figure 2C, the phase response quantifies the dispersive response of the oscillator to the changing qubit ground state, as discussed in sec.~II. The amplitude response, as shown in Fig.~2D, corresponds physically to the absorption of RF probe power by the qubit system itself. Since the measurement shown in Fig.~2 is performed with the qubits in their ground state, the gate voltage dependence of the signal amplitude is relatively weak. For qubit 1, the visibility of the gate voltage dependence is approximately $7\%$, while for qubit 2 it is below the noise level.\footnote{As in interferometry, the visiblity is defined as $\left(\mathrm{max} - \mathrm{min}\right) / \left(\mathrm{max} + \mathrm{min}\right)$.}

\begin{figure}
\centering
\includegraphics{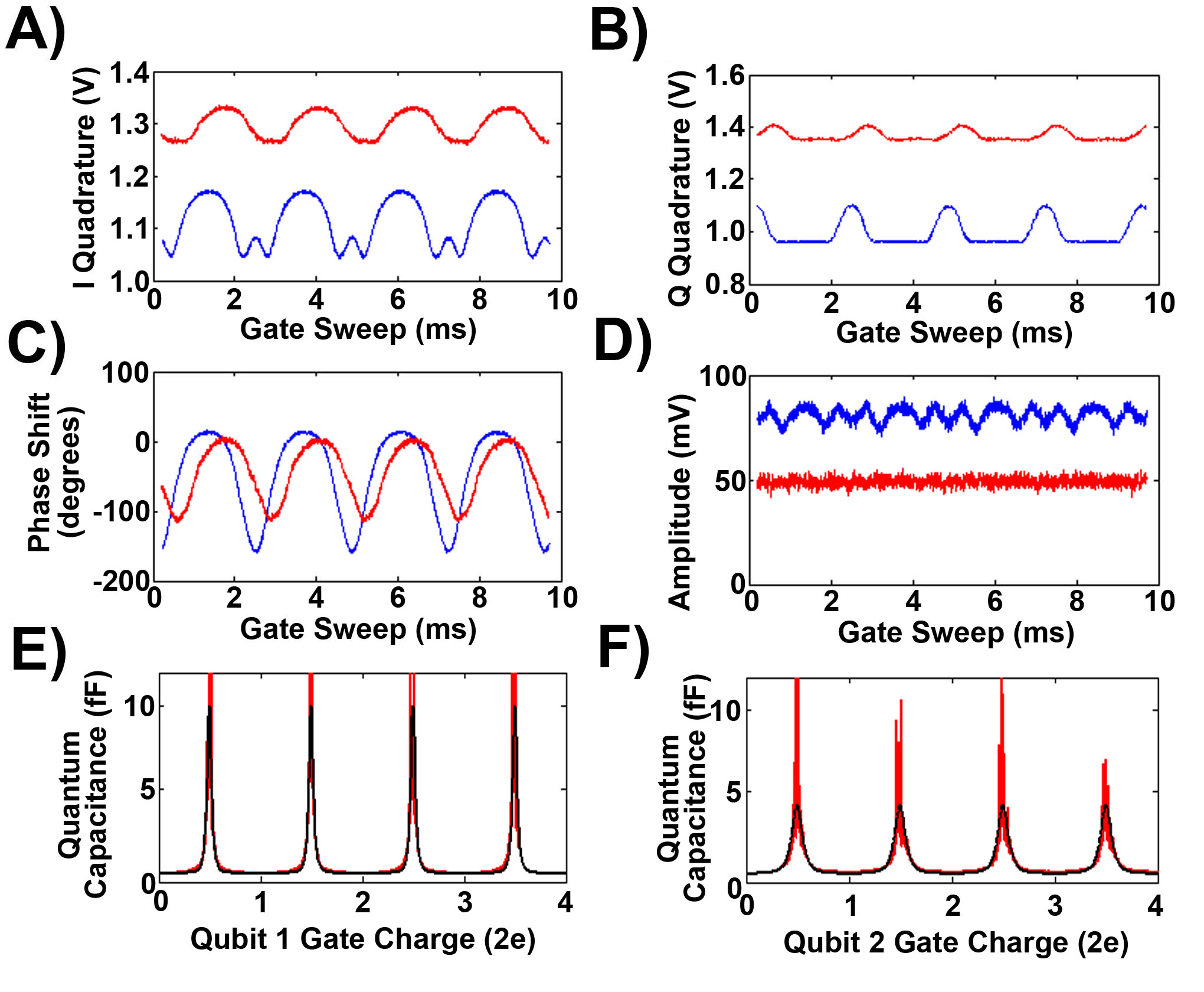}
\caption{(Color Online) Illustration of multiplexed QCM technique. Dark grey (blue) and light grey (red) curves are data sets taken simultaneously for qubits 1 and 2, respectively. \bf A)\rm~ Raw oscilloscope traces for in-phase (I) signal component. The x-axis is the sweep time of the qubit gate ramp.  \bf B)\rm~ Raw  traces for quadrature (Q) components of both qubit signals.  \bf C)\rm~ Extracted phase shift for both qubits in degrees, accounting for the loss in the tank circuit. \bf D)\rm~ Signal amplitude for both qubits. \bf E)\rm~ Quantum capacitance signal for qubit 1 extracted from the phase shift, as described in the text. The black (smooth) curves are fits to Eq.~(\ref{QConequbit}) with parameters listed in the text. \bf F)\rm~Quantum capacitance of qubit 2.}
\end{figure}

To extract the quantum capacitance from the phase shift, we use a lumped-element circuit model with the nominal parameters given in Sec.~IIb. In the case where the series resistance due to the intrinsic oscillator loss is neglected, the quantum capacitance is given by 

\begin{equation}
\label{QCinversion}
C_Q = \frac{1}{L\omega^2} \left( 1 - \frac{(\omega/\omega_C)^2}{1+Z\omega C_c} \right) - C_T
\end{equation}

\noindent where $\omega$ is the driving frequency, $\omega_c = 1/\sqrt{LC_C}$, $Z = Z_0 \left( 1 + \cos\phi(n_g) \right) / \sin\phi(n_g)$ is the overall impedance of the tank circuit, $Z_0 = 50~\Omega$ is the characteristic impedance of the transmission line, and $\phi(n_g)$ is the phase shift of the reflected signal as a function of gate voltage. This formula is used to extract the quantum capacitance as shown in Fig.~2E. The noise at the peaks is due to the fact that Eq.~(\ref{QCinversion}) is a rapidly varying function as $\phi(n_g) \rightarrow 0$, which occurs at the degeneracy point. Note that for clarity, the quantum capacitance traces for qubit 2 have been shifted horizontally by 0.5 Cooper pairs. The solid curves in Fig.~2E are theoretical plots of Eq.~(\ref{QConequbit}) for $E_{C1}/k_b=E_{C2}/k_b=190$ mK, $E_{J1}/k_b = 50$ mK, and $E_{J2}/k_b = 120$ mK. For this experiment, the flux bias was tuned to minimize $E_{J1}$. These qubit parameters are consistent with single-qubit microwave spectroscopy measurements, performed when one of the two qubits was far from its degeneracy point. The relatively small value of $E_C$ is due to the large RF gate capacitors, as shown in Fig.~1A.

\section{Coupled Qubits}
\subsection{Characterization of Coupled Qubits}

 We apply the measurement technique discussed above to characterize the coupled two-qubit system. We use two characterization techniques in two different samples. In sample 1, the qubit parameters were estimated by microwave spectroscopy. In sample 2, the qubit coupling energy was inferred by mapping the ground state capacitance as a function of both qubit gate voltages. Sample 2 had a similar design to sample 1, but with increased charging energies and an increased coupling capacitor. Using the qubit parameters for each sample, we estimate the concurrence and discuss the ground-state entanglement. Unfortunately, technical limitations prevented us from characterizing both samples using both methods. A summary of the extracted parameters is shown in Table 1. 
 
 The coupled-qubit Hamiltonian is given by Eq.~(\ref{Hamiltonian}). The coupling between the two qubits introduces avoided level crossings in the energy spectrum. The coupling also introduces a gate voltage asymmetry into the energy level diagram. By probing the system with fixed-frequency continuous-wave (cw) microwaves as a function of both gate voltages, we can map out the lowest-lying energy levels of the two-qubit system.

\begin{figure}
\centering
\includegraphics{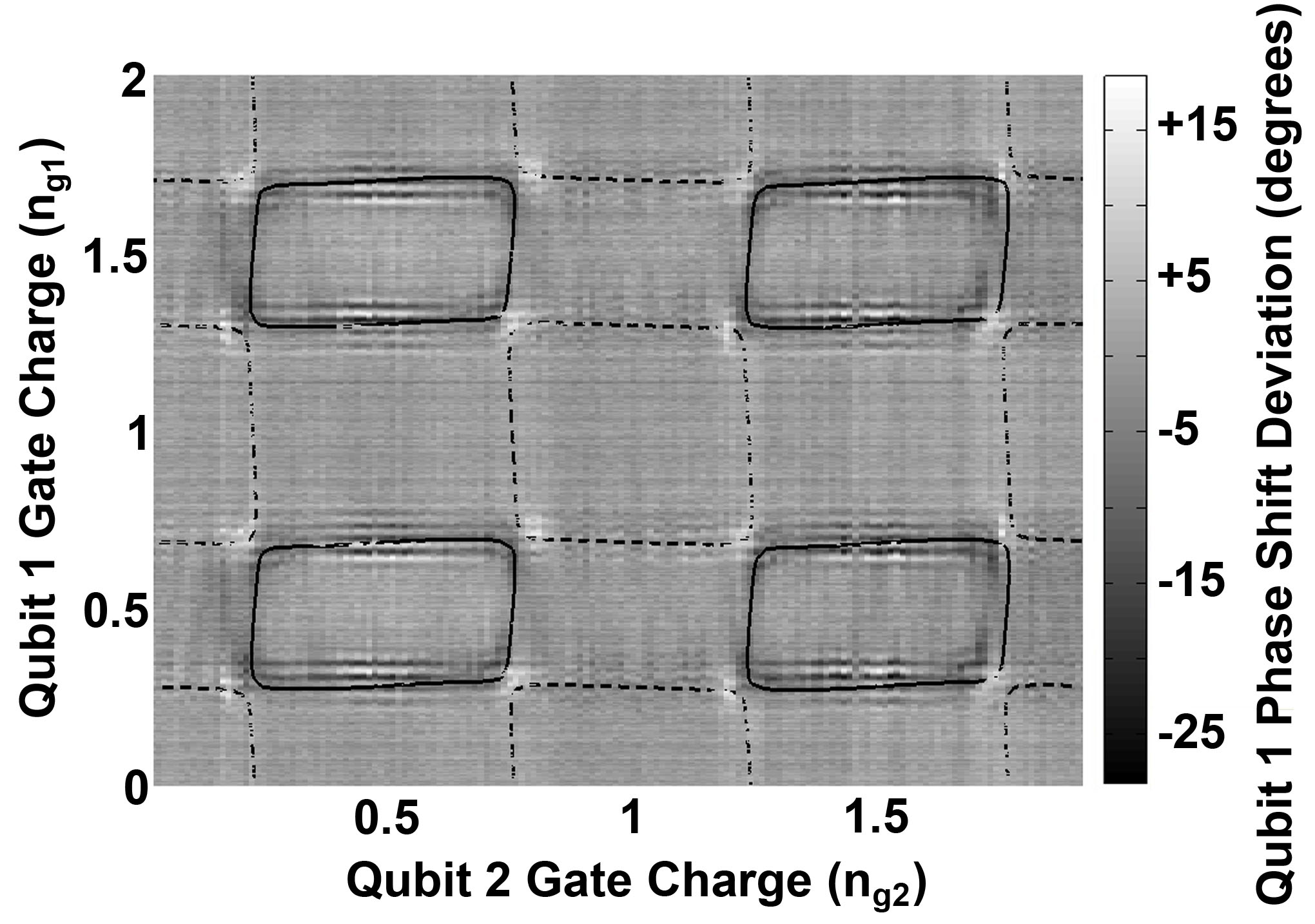}
\caption{Mesh plot of the phase shift deviation due to 11 GHz microwave excitation, as a function of both qubit gate voltages, performed using sample 1. Gate charge units are in Cooper pairs. Phase shift data taken without microwaves has been subtracted to enhance the visibility of the spectroscopic signal. Black lines are numerical calculations of energy eigenvalues commensurate with 11 GHz microwaves. Solid parallelograms correspond to transitions to the first excited state, while the dashed parallelograms correspond to transitions to the second excited state. The degree to which the parallelograms tilt is a good indicator of the coupling energy $E_m = 25 \pm 5$ mK.}
\end{figure}

	Fig.~3 shows data from such an experiment performed with sample 1, taken at a fixed microwave frequency of 11 GHz. The oscillator phase shift of qubit 1 is recorded while the system is perturbed by cw microwaves, at a range of gate voltage points for the two qubits. The phase shift signal taken without microwaves was subtracted from the data, so the plot in Fig.~3 shows the deviation in the phase shift due to the microwave excitation alone. A voltage ramp is applied to the gate of qubit 1, while the gate of qubit 2 is held at a fixed potential. After accumulating a time-averaged phase shift signal as demonstrated in Fig.~2, the gate voltage for qubit 2 is stepped to the next value. To correct for long-time DC drift in the qubit 2 gate charge offset, each averaged trace is shifted vertically to fix the location of the quantum capacitance peaks at half a Cooper pair. This also corrects for the effects of cross-capacitance between the two qubit gates. Finally, the unperturbed peaks are subtracted.  The blue represents a large negative phase deviation, indicating a more positive quantum capacitance, while the red is the reverse. The blue parallelograms correspond to transitions to the first excited state, which has a more positive quantum capacitance than the ground state. The red spots correspond to transitions to the second excited state, which has a more negative quantum capacitance. Again, in this analysis the signal is assumed to be 2e-periodic because of the observation of a transition to 1e-periodicity between 250 and 300 mK, as equilibrium quasiparticle states become occupied. 
	
	The solid lines are numerical plots of the difference between the energy levels of Eq.~(\ref{Hamiltonian}), at an energy corresponding to an 11 GHz microwave excitation. In these plots, $E_{C1}/k_b = E_{C2}/k_b = 190$ mK, $E_{J1}/k_b = 340$ mK, $E_{J2}/k_b = 430$ mK, and $E_m/k_b = 25$ mK. These values of $E_{C1,2}$ are consistent with those extracted from microwave spectroscopy performed on each qubit independently at a similar flux bias. Note that the value of the flux bias is different in this experiment compared to the measurements discussed in section IId, leading to larger values of $E_{J1,2}$. The left-leaning parallelograms indicate transitions from the ground to the first excited state, while the right-leaning parallelograms indicate transitions from the ground to the second excited state. Transitions to the third excited state are not expected at a microwave frequency of 11 GHz for these values of the qubit parameters. The coupling energy $E_m$ was extracted largely by inspecting the amount of tilt in the parallelograms, and the width of the splittings. In the absence of coupling, the parallelograms revert to rectangles. This technique conservatively permits extraction of the coupling energy within an error of $\pm 5$ mK. For sample 2, which had smaller RF gate capacitors and hence a smaller quantum capacitance, the signal-to-noise ratio was much lower and experiments of this type could not be conclusively analyzed. 
		
	We also demonstrate a second technique for establishing the coupling energy. As can be seen from Eq. (\ref{Hamiltonian}), the coupling between qubits renormalizes the effective charging energies, so that the qubit 1 charging energy ($\sigma_{z1}$ term) depends on $n_{g2}$ as well as $n_{g1}$, and vice versa for qubit 2. In a two-dimensional plot of the ground state capacitance, this leads to an electrostatic ``kink" feature in the quantum capacitance at the mutual degeneracy point. In effect, an excess charge on the island of one qubit changes the potential on the island of the other. The depth of this kink depends strongly on the interqubit coupling energy, which can be extracted by comparing the ground state capacitance as a function of both gate voltages to the theoretical values. This experiment is in some respects similar to previous ground state characterizations of three- and four-qubit systems in the flux domain.\cite{Ilichev}

\begin{table}[b]
\begin{tabular}{|l|lllllll|}
\hline
Sample & $E_{C1}$ & $E_{C2}$ & $E_{J1}$ & $E_{J2}$ & $E_m$ & Method & $\mathcal{C}$\\\hline
1 & 190 & 190 & 340 & 430 & 25 & Spectroscopy & 0.06\\
2 & 740 & 740 & 290 & 290 & 80 & Ground State & 0.27\\\hline 
\end{tabular}
\caption{Summary of extracted parameters for samples 1 and 2. All energy values are scaled by the Boltzmann constant and given in mK. The concurrence, $\mathcal{C}$, is unitless.}
\end{table}

\begin{figure}
\centering
\includegraphics{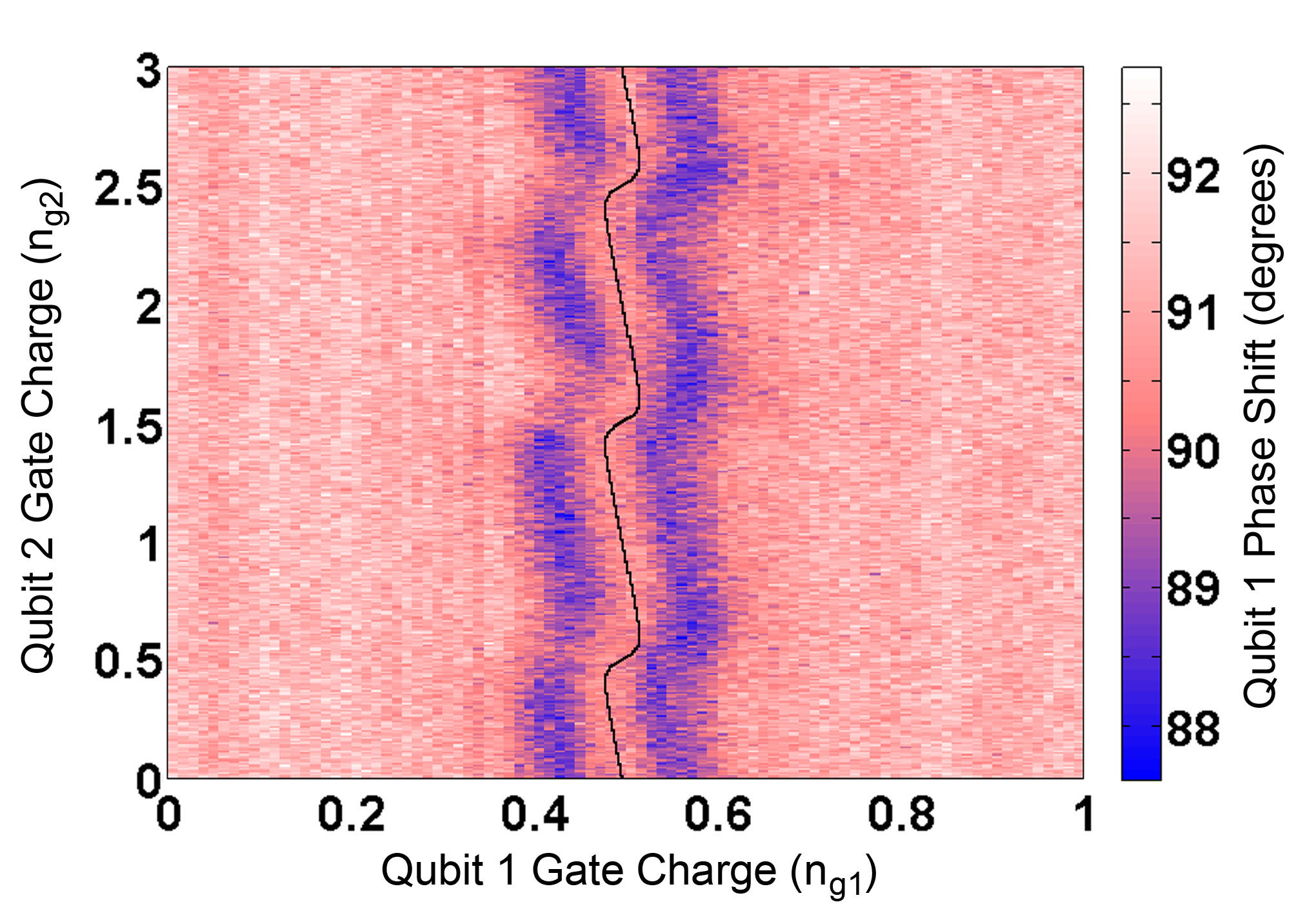}
\caption{(Color Online) Extraction of the coupling energy by fitting the ground state capacitance as a function of both gate voltages, performed using sample 2. The data is the phase shift in degrees for oscillator 1 as a function of both gate voltages. The black line is a theoretical plot of the gate voltage location of the quantum capacitance peak. From this data, we extract a coupling energy of $E_m = 80 \pm 20$ mK.}
\end{figure}

	The results of such an experiment performed using sample 2 are shown in Fig.~4. This sample had much smaller RF gate capacitors than used in sample 1, which gives a smaller phase shift signal but is less disruptive to the qubit state.  Decreasing the RF gate capacitance also increases the charging energy, which is estimated from cw spectroscopy to be $E_{C1}/k_b = E_{C2}/k_b = 740$ mK in this sample. Sample 2 also had a significantly increased coupling capacitance, so that the coupling energy $E_m$ was held to the same order of magnitude as in sample 1. The data in Fig.~4 is the phase shift in oscillator 1 as a function of both gate voltages $n_{g1}$ and $n_{g2}$.
	
	The black line in Fig.~4 is a theoretical calculation of the gate charge location of the center of the quantum capacitance peak. This calculation was done by numerically computing the ground state energy of two coupled qubits in a sixteen-level approximation (four levels for each qubit), and computing the ground state capacitance from Eq.~(\ref{capacitance}). In this simulation, the extracted parameters were $E_{C1}/k_b = E_{C2}/k_b = 740$ mK, $E_{J1} = E_{J2} = 290$ mK, and $E_m = 80$ mK. This technique is somewhat less sensitive to the parameter values than the spectroscopic technique described above, with an uncertainty in the extracted value of the coupling energy of $\pm 20$ mK. In this plot, the values of $E_{C1,2}$ and $E_{J1,2}$ are taken from cw spectroscopy measurements far from the mutual degeneracy point, so the only freely adjustable fitting parameter is the coupling energy $E_m$. This technique could not be convincingly applied to data taken with sample 1, since the qubit parameters were such that the depth of the electrostatic excursion was a much smaller fraction of the total width of the capacitance peak. As a result, it could not be reliably separated from the overall DC drift of the gate charge.

\subsection{Entanglement of Two Qubits}

In studying coupled pairs of qubits in the context of quantum information processing, it is important to quantify the degree of entanglement between the two qubits. A valuable entanglement measure for both pure and mixed states of bipartite systems is the concurrence $\mathcal{C}$, which is related in a straightforward way to the entanglement of formation.\cite{Wooters} The concurrence is an entanglement monotone which ranges from 0 for a completely separable state to 1 for a maximally entangled state. 

For a pure state $|\psi\rangle$, the concurrence of the two-qubit system is defined by 

\begin{equation}
\label{concurrence}
\mathcal{C} = \left| \langle \psi | \sigma_y \otimes \sigma_y | \psi^*\rangle \right|
\end{equation}

\noindent where $| \psi^*\rangle$ is the complex conjugate of $|\psi\rangle$. Fig.~5A shows a numerical calculation of the concurrence as a function of both gate voltages when the system is purely in the ground state. This calculation was performed using the qubit parameters of sample 2, as extracted from the data shown in Fig.~4. Note that the concurrence takes on its maximum value of 0.27 at the mutual degeneracy point, $n_{g1} = n_{g2} = 0.5$, and becomes negligible near the edges of the plot. Despite the fact that the strength of the coupling is fixed, the ground state is completely factorizable when the system is far from the mutual degeneracy point. 

\begin{figure}
\centering
\includegraphics{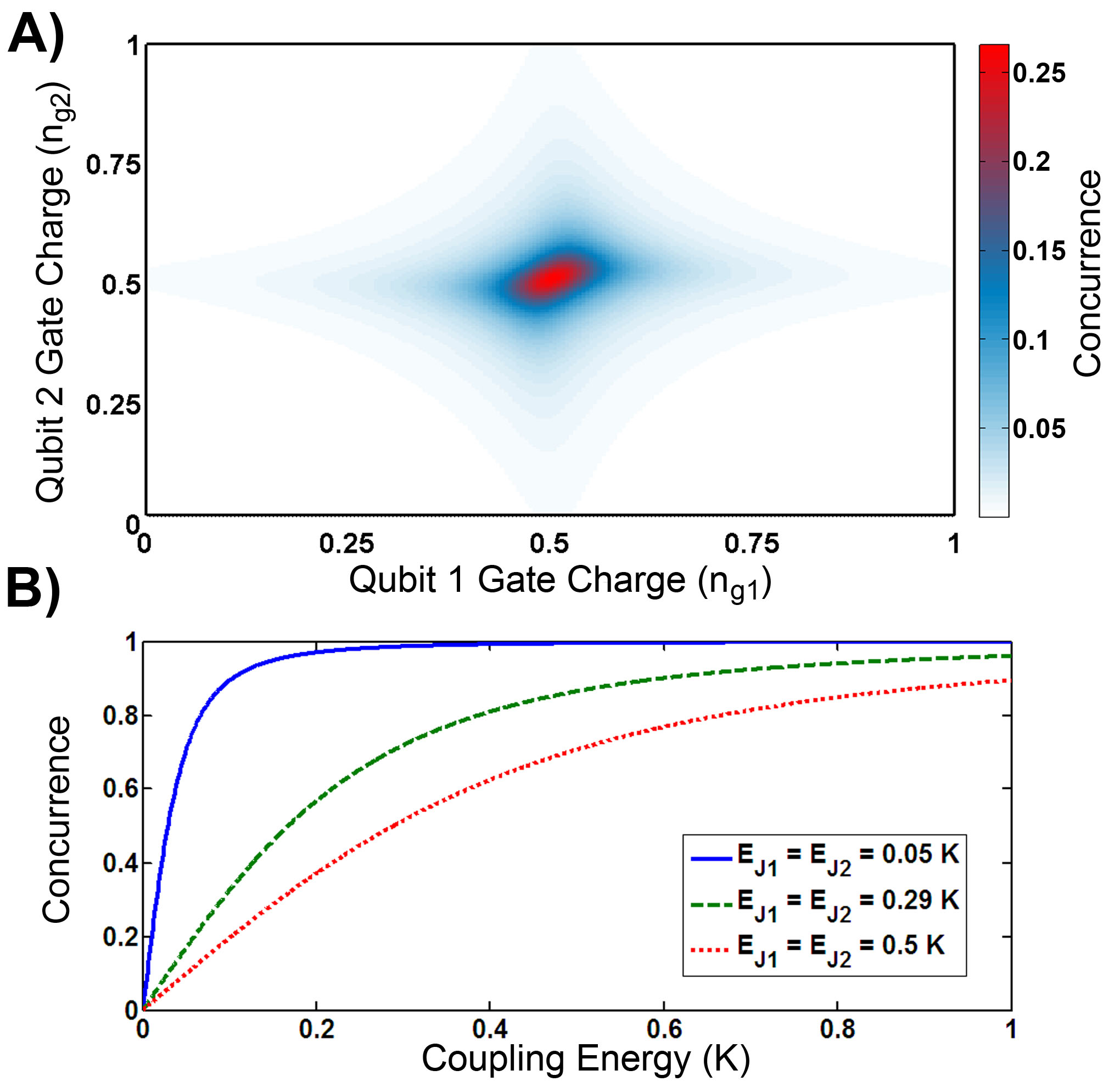}
\caption{(Color Online) Ground-state concurrence of two qubits. \bf A)\rm~The concurrence, defined by Eq.~(\ref{concurrence}) as a function of both gate voltages for the qubit parameters extracted from sample 2, as shown in Fig.~4. Here $E_{C1}/k_b = E_{C2}/k_b = 740$ mK, $E_{J1}/k_b = E_{J2}/k_b = 290$ mK, and $E_m/k_B = 80$ mK. \bf B)\rm~Concurrence at the mutual degeneracy point in the special case of two identical qubits, plotted as a function of coupling energy for three different values of $E_J$. The green (dashed) curve shares the same parameter values as the plot in (A).}
\end{figure}

	In the particular case of identical qubits, where $E_{C1} = E_{C2}$ and $E_{J1} = E_{J2}$, the problem simplifies considerably. When expressed in the singlet-triplet basis, the singlet and triplet subspaces of Hamiltonian (\ref{Hamiltonian}) are fully decoupled, and the Hamiltonian can be cleanly diagonalized in closed form.\cite{Storcz} At the degeneracy point, the ground state of the two-qubit system is given by 
	
\begin{equation}
\label{groundstate}
|g\rangle = \frac{1}{\sqrt{2(A^2 + 1)}} \left[ |00\rangle + |11\rangle + A \left(|01\rangle + |10\rangle\right) \right]
\end{equation}

\noindent where $A = \left( E_m + \sqrt{E_m^2 + E_J^2} \right) / E_J$. This leads to a simple expression for the concurrence, $\mathcal{C} = \left| \frac{A^2 - 1}{A^2 + 1} \right|$, in the zero-temperature limit.  Note from this formula that $\mathcal{C} \rightarrow 0$ as $E_m \rightarrow 0$, and $\mathcal{C} \rightarrow 1$ as $E_m \rightarrow\infty$.  This expression is plotted in Fig.~5B as a function of coupling energy, for three different values of $E_J$. Note that the green (dashed) curve corresponds to the qubit parameters used to construct Fig.~5A, $E_{J1}/k_b = E_{J2}/k_b = 200$ mK. The blue (solid) and red (dotted) curves correspond to $E_{J1}/k_b = E_{J2}/k_b = 50$ mK and $E_{J1}/k_b = E_{J2}/k_b = 400$ mK, respectively. Using the qubit parameters extracted in section II, we find that the concurrence at the mutual degeneracy point $n_{g1} = n_{g2} = \frac{1}{2}$ is $\mathcal{C} = 0.06$ for sample 1 and $\mathcal{C} = 0.27$ for sample 2. Note that at the mutual degeneracy point, the concurrence does not depend on $E_{c1,2}$, although it is strongly dependent on $E_{J1,2}$ and $E_m$. 

\section{Conculsions}

We have used a multiplexed quantum capacitance measurement technique to characterize a system of two entangled superconducting qubits. We have determined the energy scales of the two-qubit system, both through microwave spectroscopy and an examination of the ground state capacitance. From this information, we have estimated the concurrence, the degree of entanglement between the two qubits, and discussed the ground state entanglement for superconducting qubits. 

 In the multiplexed QCM technique, two on-chip superconducting lumped-element LC oscillators are capacitively coupled to the qubit islands, and monitored thorough a single RF line. Since the overall capacitance is dependent on the qubit state, a measurement of the oscillator with RF reflectometry constitutes a dispersive measurement of the qubit, which is applicable directly at the degeneracy point. This technique is readily scalable to read out a large array of qubits, which is a key ingredient in the development of a large-scale quantum computer. 

\acknowledgements{We would like to thank Richard Muller for performing the electron-beam lithography. This work was conducted at the Jet Propulsion Laboratory, California Institute of Technology, under a contract with the National Aeronautics and Space Administration (NASA), and was funded by a grant from the National Security Agency.  Juan Bueno acknowledges support from NASA. Matthew Shaw acknowleges support from the University of Southern California College of Arts and Sciences. }

\end{document}